\documentclass{PoS}
\usepackage{amsmath}	

\newcommand{\ket}[1]{\left| #1 \right>}
\newcommand{\braket}[2]{\left< #1 | #2 \right>}
\newcommand{\Braket}[3]{\left< #1 \left| #2 \right| #3 \right>}

\title{Stringy excitation and role of UV gluons\\
       in lattice QCD }

\ShortTitle{Stringy excitation and role of UV gluons in lattice QCD }

\author{\speaker{Hiroshi Ueda}, Takahiro M. Doi, Sho Fujibayashi, Shoichiro
        Tsutsui, Takumi Iritani, Hideo Suganuma \\ 
        Department of Physics, 
        Kyoto University, 
        Kitashirakawaoiwake, Sakyo, Kyoto 606-8502, Japan\\
        E-mail: \email{ueda@ruby.scphys.kyoto-u.ac.jp}}

\abstract{Using SU(3) quenched lattice QCD, 
we study ground-state and low-lying even-parity 
excited-state potentials of quark-antiquark systems 
in terms of the gluon-momentum component in the Coulomb gauge.
By introducing UV-cut in the gluon-momentum space, 
we investigate the ``UV-gluon sensitivity'' 
of the ground-state and excited-state potentials 
and the stringy excitation quantitatively.
Even after cutting off high-momentum gluon component above 1.5GeV, 
the IR part of the ground-state potential is almost unchanged. 
On the other hand, 
the change of excited-state potential is more significant by 
the cut of UV-gluons. 
However, even after the removal of UV-gluons, the magnitude of 
the low-lying gluonic excitation remains to be of the order of 1GeV. 
}

\FullConference{
The 30 International Symposium on Lattice Field Theory - Lattice 2012,\\
		June 24-29, 2012\\
		Cairns, Australia}

\begin{document}

\section{Introduction}

Unlike QED, quantum chromodynamics (QCD) forms a color-electric flux-tube 
between the quark and the antiquark in mesonic systems, 
and this one-dimensional squeezing of the color-electric 
field leads to a linear confinement potential 
in the infrared region \cite{Nambu:1974}. 
Actually, apart from the color-Coulomb energy around quarks, 
the flux-tube formation has been observed in lattice QCD 
both for $Q \bar Q$ \cite{Bali:1994de} and 3$Q$ systems 
\cite{Ichie:2002dy,Takahashi:2003,Bornyakov:2004}. 

In the flux-tube picture of hadrons, which is idealized as
the string picture in the infrared region, one can expect 
``stringy excitations'' of hadrons, as shown in Fig.1. 
This stringy mode is non-quark-origin excitation, 
and therefore it can be regarded as a gluonic excitation. 
Such a gluonic-excited state would be interpreted as 
hybrid hadrons ($q \bar qG$ and $qqqG$), which are interesting hadrons 
beyond the quark-model framework \cite{Takahashi:2003}. 

\begin{figure}[h]
\begin{center}
\includegraphics[scale=0.7]{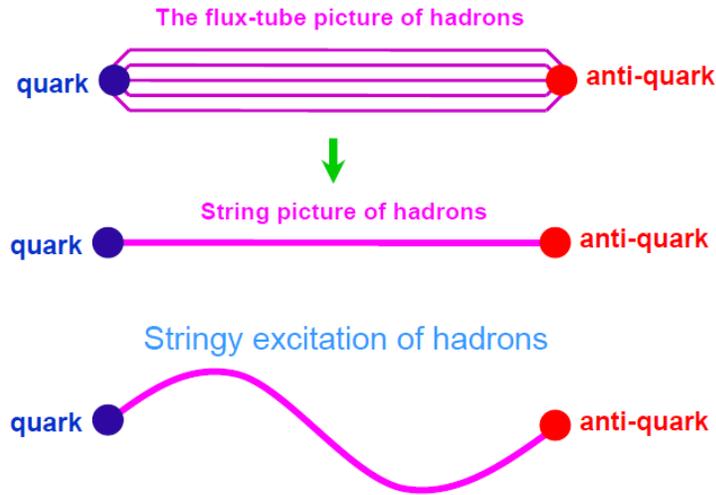}
\caption{
Schematic illustration of the stringy excitation of hadrons. 
The flux-tube picture of hadrons is idealized as 
the string picture in the infrared region, 
which is expected to allow ``stringy excitations'' of hadrons. 
Since this stringy mode is non-quark-origin excitation, 
it can be regarded as a gluonic excitation. 
}
\end{center}
\end{figure}

In lattice QCD, from a detailed Wilson-loop analysis, 
the excited-state potentials and the gluonic excitation 
have been calculated both for spatially-fixed 
$Q\bar Q$ systems \cite{Juge:2002br} 
and for $3Q$ systems \cite{Takahashi:2002it,Takahashi:2004rw}. 
For simpler $Q\bar Q$ systems, the behavior of the gluonic excitation 
is almost consistent with the string excitation in 
infrared region, in spite of a significant difference 
at the small distance \cite{Juge:2002br}. 

In the previous work, IR/UV-gluon contribution to 
the ground-state potential has been studied 
\cite{Yamamoto:2008am,Yamamoto:2008ze}, and 
the confinement force is found to be almost unchanged 
even after the cut of high-momentum gluon components 
above 1.5GeV \cite{Yamamoto:2008am,Yamamoto:2008ze} in the Landau gauge.
This means that the confinement property is insensitive to UV gluons.

In this work, we study not only ground-state potential but also 
low-lying even-parity excited-state potentials of $Q\bar Q$ systems 
in terms of gluon momentum component in the Coulomb gauge. 
By introducing UV-cut in three-dimensional gluon-momentum space, 
we study the UV-gluon contribution to 
excited-state potentials and stringy excitations. 

\section{Lattice formulation}

\subsection{Formalism to extract excited-state potentials 
in lattice QCD}

We present the formalism to extract the excited-state potential 
\cite{Takahashi:2002it,Takahashi:2004rw} 
for the spatially-fixed $Q\bar Q$ system. 
We denote the $n$th eigen-state of the QCD Hamiltonian $H$ by $\ket{n}$, 
\begin{eqnarray}
 H\ket{n} &=& V_n\ket{n}. 
\end{eqnarray}
Here, $V_n$ denotes $n$th excited-state potential, 
and $0$th eigen-state means the ground-state. 
Consider arbitrary independent $Q\bar Q$ states $\ket{\phi_k}(k=0,1,2...)$. 
Generally, each $Q \bar Q$ state $\ket{\phi_k}$ can be expressed by 
a linear combination of the $Q\bar Q$ physical eigen-states: 
\begin{eqnarray}
 \ket{\phi_k}= c^k_0\ket{0} + c^k_1\ket{1} + c^k_2\ket{2} + \dots~ .
\end{eqnarray}

The Euclidean-time evolution of the $Q\bar Q$ state $\ket{\phi_k(t)}$ is 
expressed with the operator $e^{-Ht}$, which corresponds to 
the transfer matrix in lattice QCD. 
The overlap $\braket{\phi_j(T)}{\phi_k(0)}$ is given by 
the Wilson loop $W_T^{jk}$, 
sandwiched by initial state $\phi_k$ at $t =0$ 
and final state $\phi_j$ at $t =T$, 
and is expressed in the Euclidean Heisenberg picture as 
\begin{eqnarray}
 W_T^{jk} &\equiv& \braket{\phi_j(T)}{\phi_k(0)} =
  \Braket{\phi_j}{W(T)}{\phi_k} = \Braket{\phi_j}{e^{-HT}}{\phi_k}\\
  &=& \sum_{m=0}^\infty \sum_{n=0}^\infty \bar c_m^j c_n^k
  \Braket{m}{e^{-HT}}{n} = \sum_{n=0}^\infty \bar c_n^j e^{-V_nT} c_n^k,
\end{eqnarray}
with the complex-conjugate notation of $\bar c_n^j \equiv (c_n^j)^*$. 
This is a basic relation between Wilson loops and potentials. 
By introducing the matrices $C$ and $\Lambda_T$ such that 
\begin{eqnarray}
C^{nk} = c_n^k, \qquad \Lambda_T^{mn}=e^{-V_nT} \delta^{mn}, 
\end{eqnarray}
this relation can be rewritten as 
\begin{eqnarray}
 W_T = C^\dagger \Lambda_T C. 
\end{eqnarray}
In general, $C$ is not a unitary matrix, 
and depends on the choice of $\ket{\phi_k}$. 
Using this relation, we extract the potentials $V_n~(n=0,1,2 \cdots)$ 
from the Wilson loop $W_T$. Consider the following combination: 
\begin{eqnarray}
 W_T^{-1}W_{T+1} = \left\{ C^\dagger \Lambda_T C \right\}^{-1} C^\dagger
  \Lambda_{T+1} C = C^{-1}{\rm diag}\left( e^{-V_0}, e^{-V_1}, e^{-V_2},
				    \dots\right) C.
\end{eqnarray}
Then, $e^{-V_n}$ can be obtained as the eigen-values of the matrix 
$W_T^{-1}W_{T+1}$. 
In fact, they are the solutions of the secular equation, 
\begin{eqnarray}
{\rm det}\left\{ W_T^{-1}W_{T+1} -t {\bf 1} \right\} 
  = \prod_n \left( e^{-V_n} -t \right) =0.
\label{Eq:secular}
\end{eqnarray}
In this way, the potentials $V_n~(n=0,1,2,...)$ can be obtained 
from the Wilson loop matrix, $W_T^{-1}W_{T +1}$. 

In the practical calculation, we prepare gauge-invariant $Q \bar Q$ 
states $\ket{\phi_k}$ composed by fat-links obtained with APE smearing method 
\cite{Albanese:1987ds}, 
and calculate many Wilson loops sandwiched by various combination 
of initial state $\ket{\phi_k}$ and final state $\ket{\phi_j}$ . 
By solving the secular equation Eq.~\eqref{Eq:secular} 
within a truncated dimension, 
ground-state and excited-state potentials can be obtained. 

\subsection{Discrete Fourier transformation 
and UV-cut of gluon momentum components}

In this subsection, we consider 
the three-dimensional Fourier transformation of 
the link-variable $U_\mu(x) \in {\rm SU(3)}$ on 
a periodic lattice of size $L^4$, 
and introduce UV-cut in three-dimensional momentum space 
\cite{Yamamoto:2008am,Yamamoto:2008ze}. 
For the argument on the gluon momentum, 
gauge fixing is generally needed. 
For the comparison with continuum QCD, 
the suitable gauge to be taken on lattice would be 
the Landau or the Coulomb gauge, 
where the gauge field tends to be continuous. 

Here, we consider link-variables fixed in the Coulomb gauge, 
because spatial gauge-field fluctuation is strongly suppressed. 
The Coulomb gauge has a global definition to minimize the ``total amount 
of the spatial gauge-field fluctuation'', 
\begin{eqnarray}
R \equiv \int\!\! d^3\! x~ {\rm tr} \left\{A_i(\vec x,t)A_i(\vec x,t)\right\}
=\frac{1}{2} \int\!\! d^3\! x~ A_i^a(\vec x,t)A_i^a(\vec x,t)
\end{eqnarray}
The Coulomb gauge has a physical meaning that 
it maximally suppresses artificial fluctuation 
stemming from gauge degrees of freedom for spatial gluons. 
In lattice QCD, the Coulomb gauge fixing is expressed in terms of 
link-variable and is defined by the maximization of 
\begin{eqnarray}
R_{\rm latt} \equiv \sum_{\vec x} \sum_i {\rm Re~tr~}U_i(\vec x,t).
\end{eqnarray}

\begin{figure}[h]
\begin{center}
\includegraphics[scale=0.4]{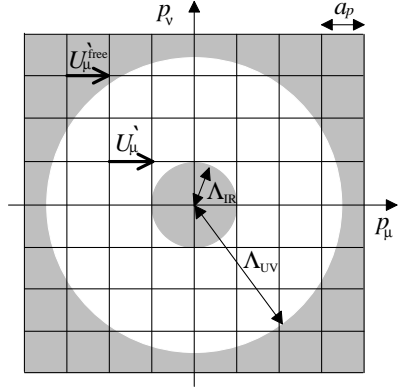}
\caption{
A schematic figure of UV-cut with $\Lambda_{\rm UV}$ 
and IR-cut with $\Lambda_{\rm IR}$ on momentum-space lattice, 
of which spacing is given by $a_p\equiv 2\pi/(La)$. 
The momentum-space link-variable ${\tilde U}_{\mu}(p)$ is replaced 
by the free-field link-variable 
${\tilde U}^{\rm free}_{\mu}(p)=\delta_{p0}$ 
in the shaded cut regions. 
}
\end{center}
\end{figure}

Now, we perform the three-dimensional discrete Fourier transformation 
of the link-variable $U_\mu(x) \in {\rm SU(3)}$, 
and define the ``momentum-space link-variable'': 
\begin{eqnarray}
 \tilde{U}_\mu(\vec p,t) \equiv \frac{1}{L^3}\sum_{\vec x} 
U_\mu(\vec x,t)\exp{\left(i\vec p\cdot \vec x\right)}.
\end{eqnarray}
We introduce ``UV-cut'' in the momentum space. 
Outside the cut, $\tilde U_\mu(\vec p,t)$ is replaced 
by 0, since momentum-space link-variable is 
${\tilde U}^{\rm free}_{\mu}(p)=\delta_{p0}$ 
in the free-field case of $U_\mu^{\rm free}(x) \equiv 1$. 
(See Fig.1.) 
We define ``UV-cut momentum-space link-variable'': 
\begin{eqnarray}
 \tilde U_\mu^\Lambda(\vec p,t) \equiv 
 \begin{cases}
 \tilde U_\mu(\vec p,t) & 
~~~~{\rm for}~~~~~~ |\vec p| \le \Lambda_{\rm UV} \\
            0           & 
~~~~{\rm for}~~~~~~ |\vec p| >   \Lambda_{\rm UV} 
 \end{cases}
\end{eqnarray}

By the three-dimensional inverse Fourier transformation 
\begin{eqnarray}
 U'_\mu(\vec x,t) \equiv 
    \sum_{\vec p} \tilde U_\mu^\Lambda(\vec p,t) 
    \exp{\left(-i\vec p \cdot \vec x \right)},
\end{eqnarray}
and SU(3) projection by maximizing 
\begin{eqnarray}
{\rm Re~tr}\left\{U_\mu^{\Lambda}(\vec x,t) U_\mu'^{\dagger}(\vec x,t)\right\},
\end{eqnarray}
we obtain ``UV-cut (coordinate-space) link-variable'': 
\begin{eqnarray}
 U^\Lambda_\mu (\vec x,t) \in {\rm SU(3)}. 
\end{eqnarray}
Using the UV-cut link-variable $U^\Lambda_\mu (x)$ instead of $U_\mu (x)$, 
we calculate many Wilson loops $W_T^{ik}$ sandwiched by various 
combination of initial state $\ket{\phi_k}$ and final state 
$\ket{\phi_j}$
Note here that the UV-cut should be introduced also to $U_4(x)$. 
Otherwise, the QCD Hamiltonian is not changed, so that 
the potentials $V_n$ are not changed at all.

\section{Ground-state and excited-state $Q\bar Q$ potentials 
and gluonic excitation energy}

In this section, we show the lattice QCD results of 
the ground-state/excited-state potentials and gluonic excitation energy 
in $Q \bar Q$ systems with or without UV-cut.
Our numerical simulation is performed with isotropic plaquette gauge action 
with $\beta = 6.0$ at the quenched level. 
The lattice size is $16^4$, and the periodic boundary condition is imposed. 
This lattice QCD condition corresponds to 
the (coordinate-space) lattice spacing $a\simeq$ 0.104fm, and 
the momentum-space lattice spacing $a_p \equiv 2\pi/(La) \simeq$ 0.74GeV. 
We use 100 gauge configurations, 
and average all the parallel-translated Wilson loops in each configuration.

\begin{figure}[h]
 \begin{center}
  \begin{minipage}{0.32\hsize}
   \includegraphics[width=\hsize]{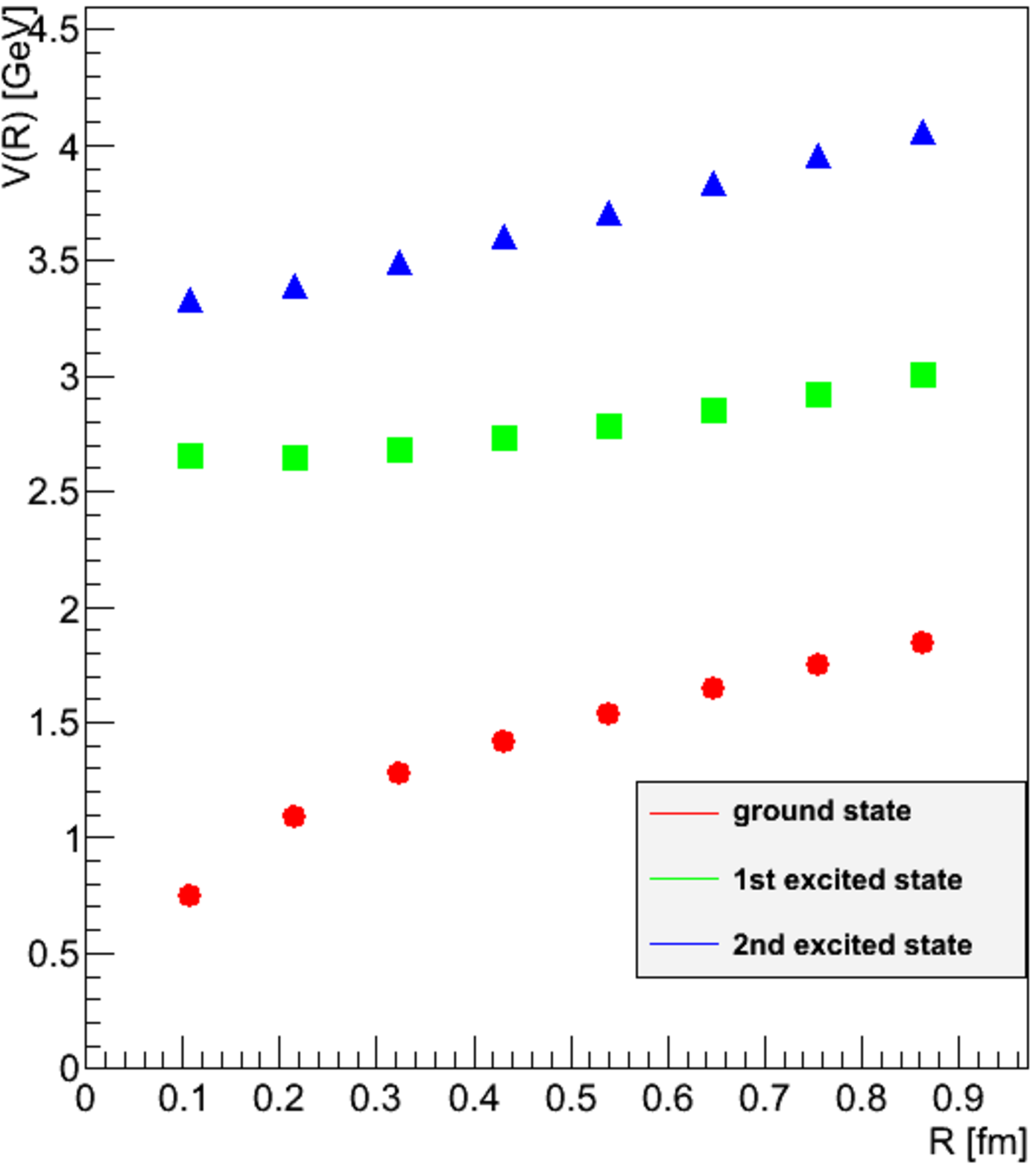}  
  \end{minipage}
  \begin{minipage}{0.32\hsize}
   \includegraphics[width=\hsize]{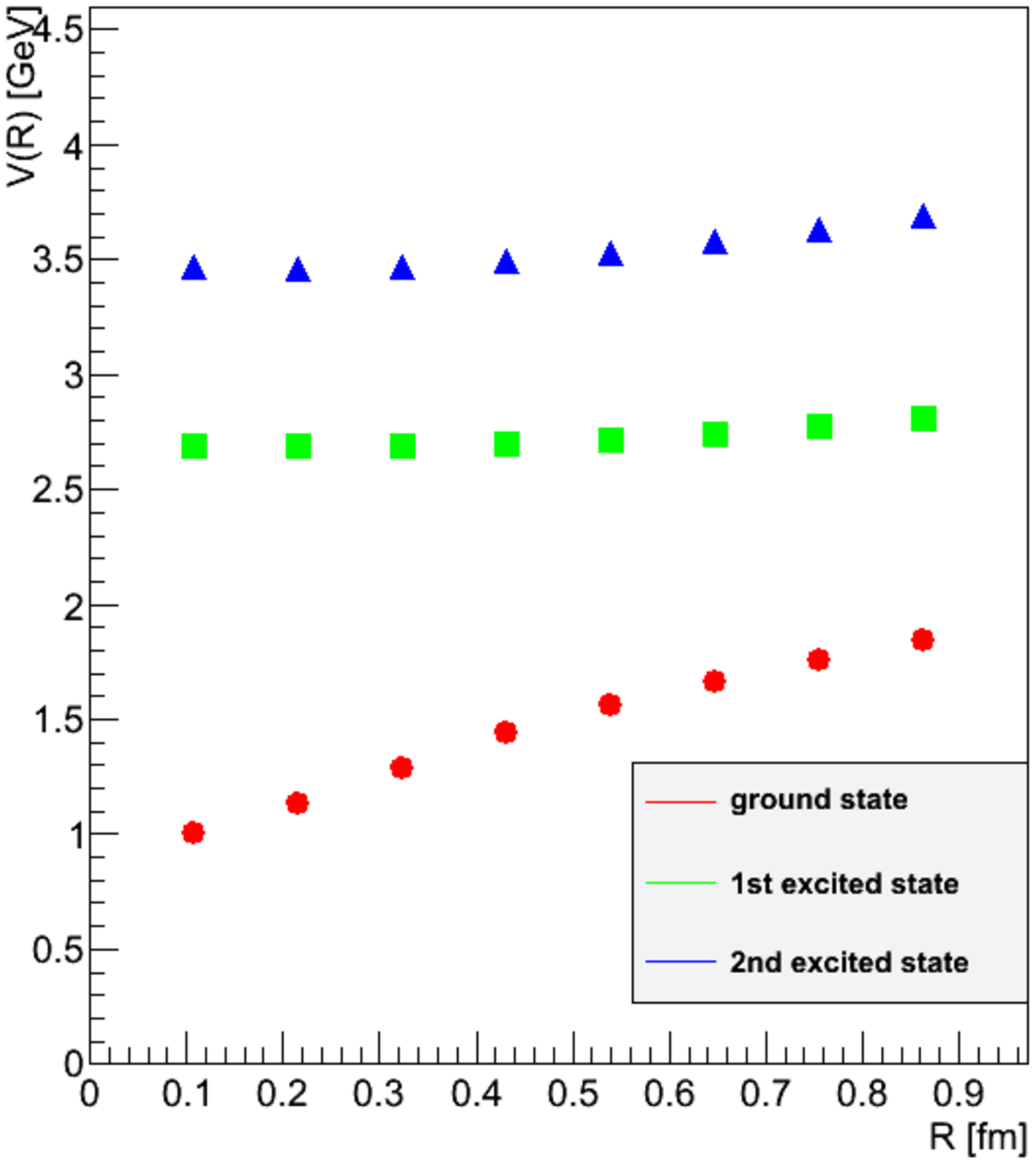}  
  \end{minipage}
  \begin{minipage}{0.32\hsize}
   \includegraphics[width=\hsize]{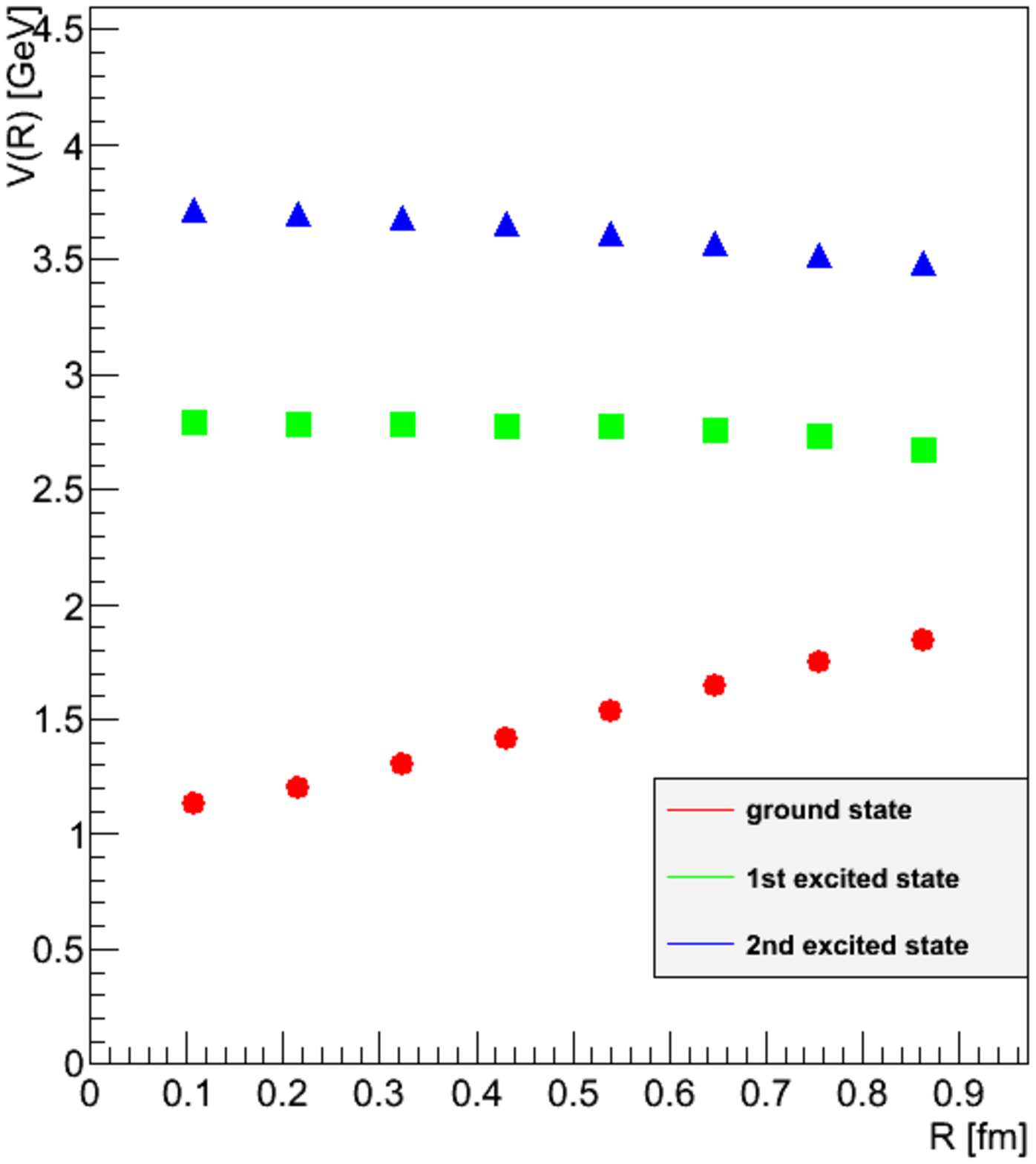}  
  \end{minipage}
 \end{center}
\caption{Ground-sate and even-parity excited-state potentials in 
 $Q \bar Q$ systems with or without UV-cut plotted against 
 the interquark distance $R$. 
 The left panel shows the result without UV-cut. 
 The middle and right panels show the results with the UV-cut of 
 $\Lambda_{\rm UV}=3a_p \simeq 2.2{\rm GeV}$ and 
 $\Lambda_{\rm UV}=2a_p \simeq 1.5{\rm GeV}$, respectively. 
 The circle symbol denotes the ground-state potential. 
 The square and the triangle symbols denote 
 the even-parity excited-state potentials.}
\label{Fig:QPOT}
\end{figure}

For simplicity, we only consider even-parity exited-state potentials 
in this paper. 
We prepare the $Q \bar Q$ state
$\ket{\phi_k}(k=0,1,2,3)$ composed by the ``fat-links'' obtained 
by the APE smearing method \cite{Albanese:1987ds} with 
the smearing parameter $\alpha = 2.3$ and the iteration number of 
$N_{\rm smr}=0, 8, 16, 24$.
Note that only even-parity components can be obtained 
by this parity-invariant procedure. 

\subsection{Ground-state and excited-state $Q\bar Q$ potentials with UV cut}

Figure~\ref{Fig:QPOT} shows ground-state and excited-state potentials 
in $Q \bar Q$ systems with or without UV-cut of gluon fields. 
In the original no UV-cut case, the IR slopes of ground-state 
and excited-state potentials are almost the same, 
as was indicated by the previous lattice studies \cite{Juge:2002br}. 
This means the same confinement force in the infrared region. 

By the cut of UV-gluon above $\Lambda_{\rm UV} = 3a_p \simeq 2.2$GeV, 
the short-distance Coulomb part proportional to $1/r$ 
reduces in ground-state potential. 
In the case of $\Lambda_{\rm UV} = 2a_p \simeq 1.5$GeV, 
the short-distance Coulomb part disappears in the ground-state potential. 
These tendencies are consistent with the previous studies 
\cite{Yamamoto:2008am,Yamamoto:2008ze}.

On the other hand, 
the shape of the excited-state potential is largely changed 
by the UV-cut of gluon fields for $\Lambda_{\rm UV}=1.5,~2.2{\rm GeV}$, 
while the ground-state potential 
is not so changed except for the short distance.
As a caution, the physical size of our lattice is rather small, 
and the true IR slope of the excited-state is expected to be unchanged, 
because no change is found in the ground-state potential, 
which gives a lower bound of $V_n$ in the infrared region. 
In any case, the change of the excited-potential is more significant 
than that of the ground-state potential by the UV-cut of gluons. 

In the string picture, this result seems to be natural as mentioned below. 
For the stringy-excited state as shown in Fig.1, 
there is a typical wavelength proportional to the interquark distance $R$, 
and this wavelength is smaller for higher excitation mode. 
Then, we expect a significant influence of the removal of UV-gluons 
for the stringy mode, 
when the UV-cut length $1/\Lambda_{\rm UV}$ becomes larger than 
the typical wavelength of the stringy excitation. 
In fact, the effect of UV-gluon cut would be larger for higher excitation. 
Our lattice QCD results seem to be qualitatively consistent with this tendency.

\subsection{Gluonic excitation energy with UV cut}

\begin{figure}[h]
 \begin{center}
  \begin{minipage}{0.32\hsize}
   \includegraphics[width=\hsize]{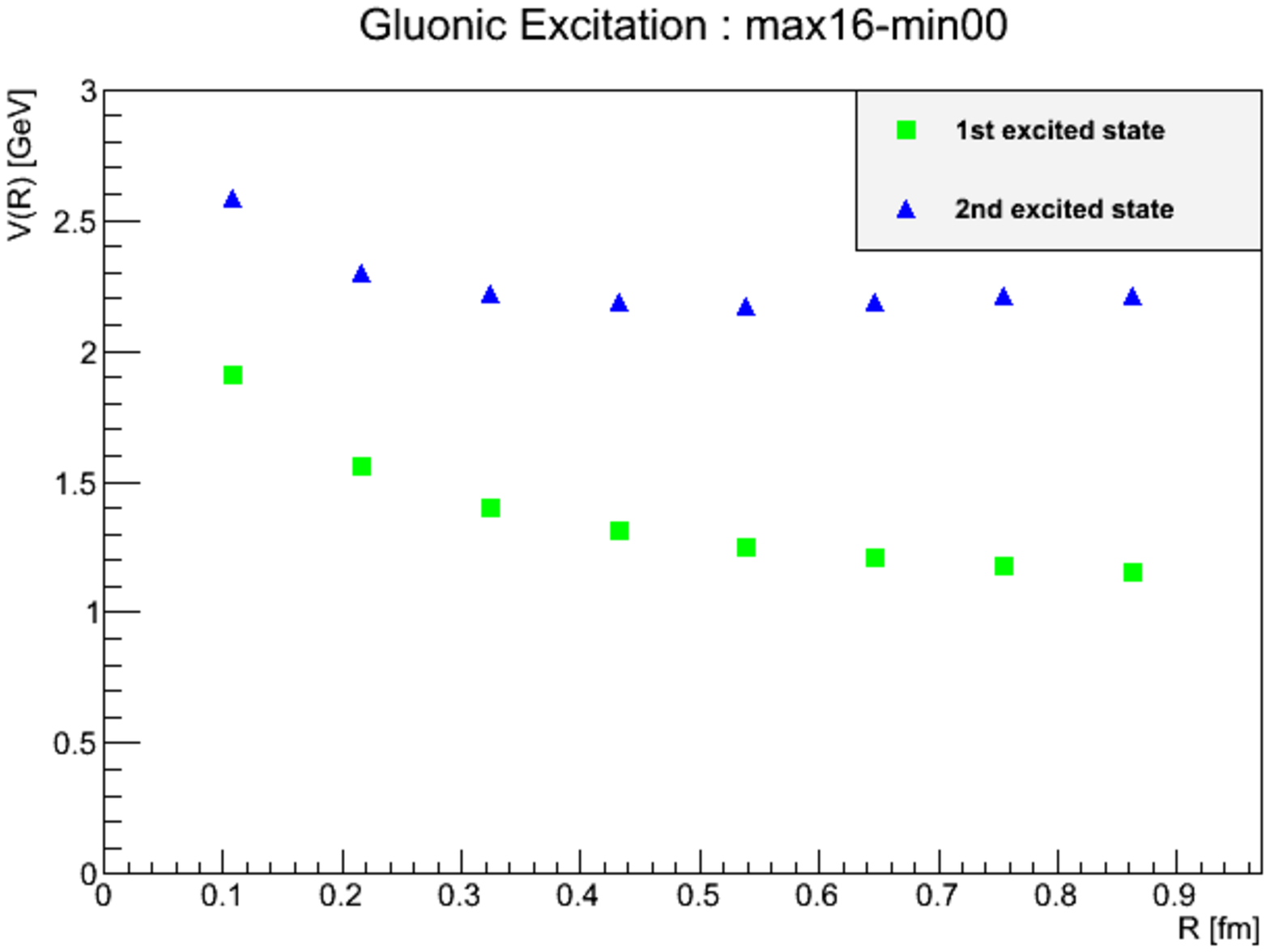}
  \end{minipage}
  \begin{minipage}{0.32\hsize}
   \includegraphics[width=\hsize]{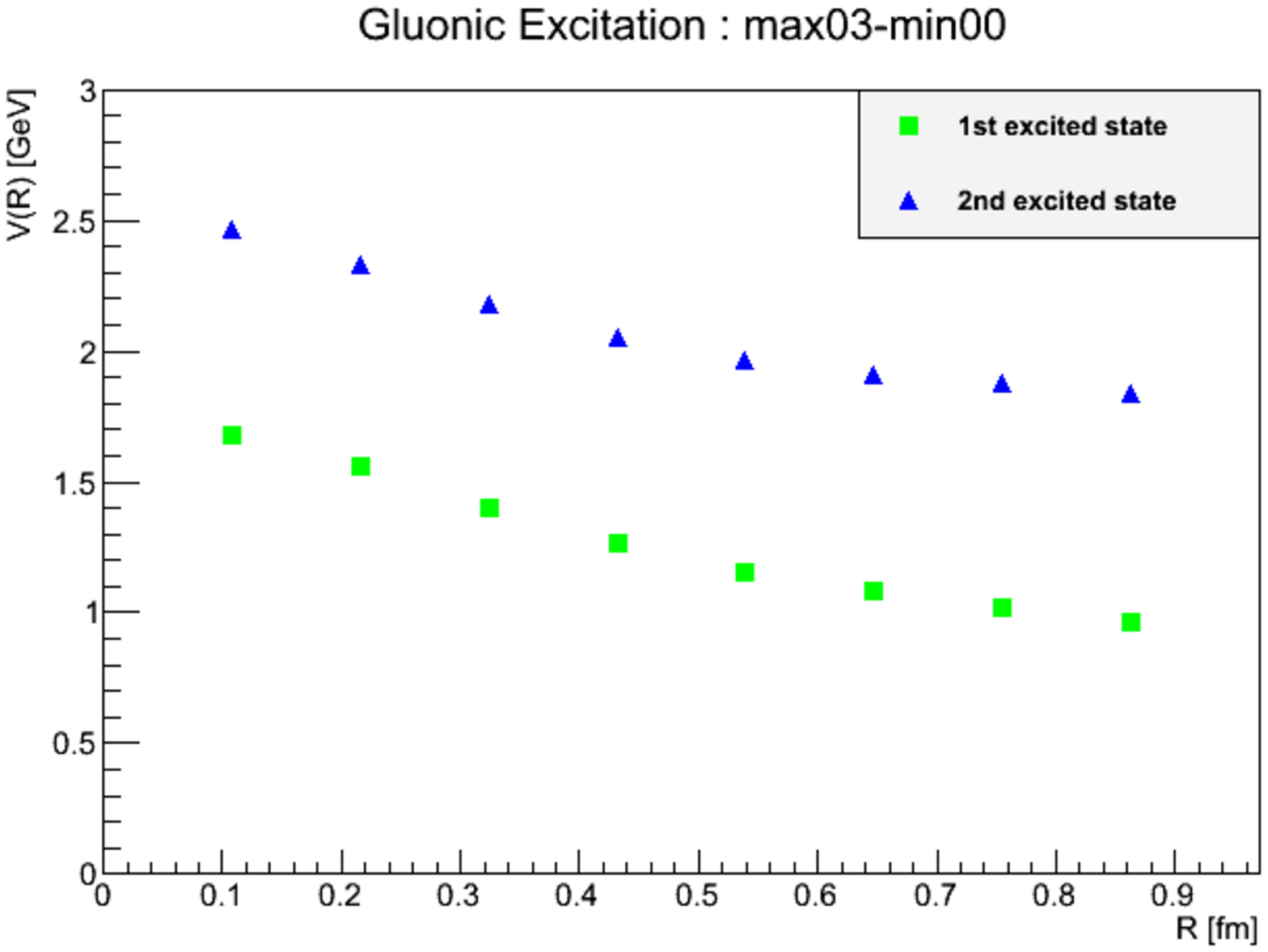}
  \end{minipage}
  \begin{minipage}{0.32\hsize}
   \includegraphics[width=\hsize]{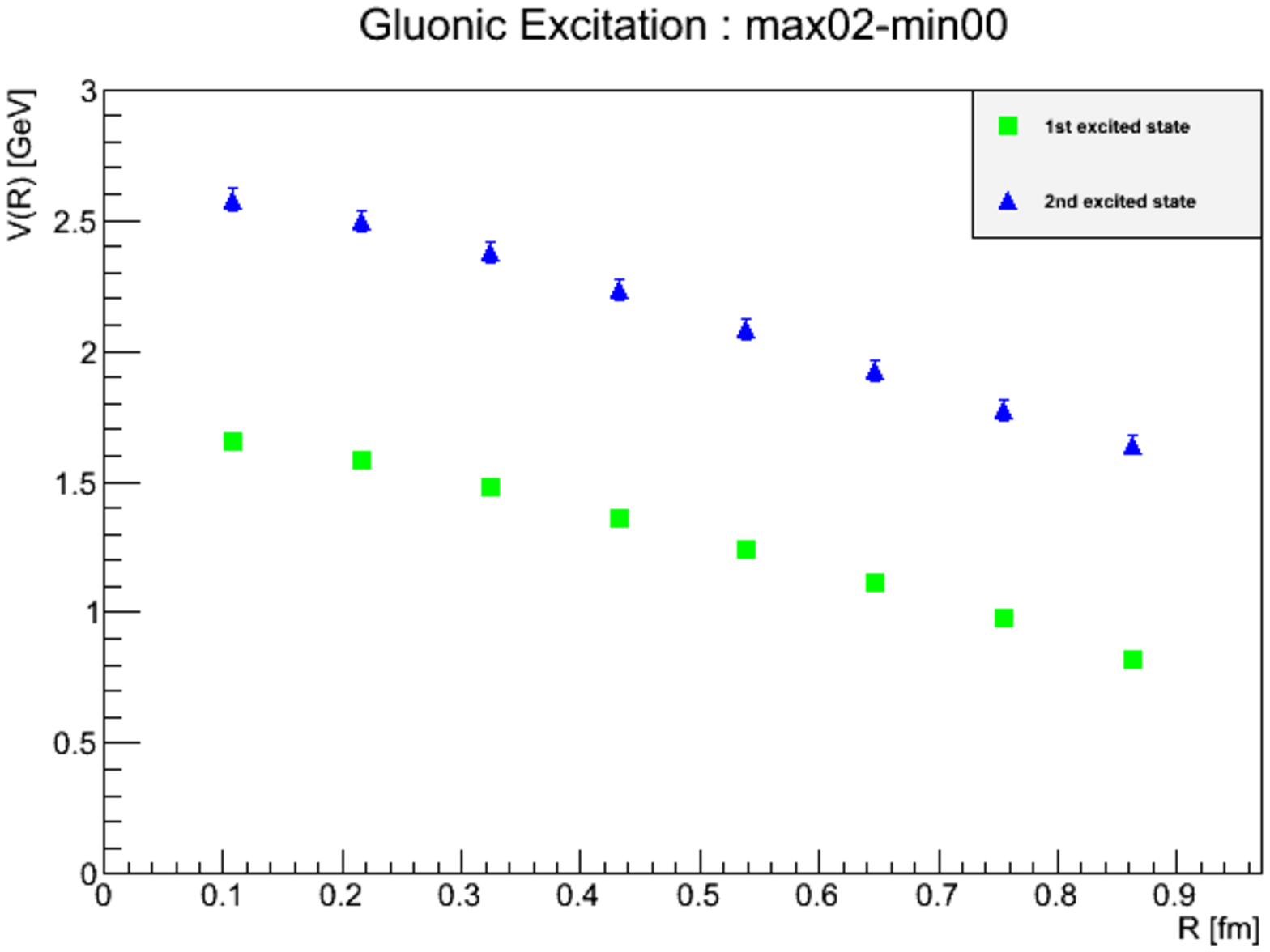}
  \end{minipage}
 \end{center}
 \caption{The gluonic excitation energy defined by 
the difference between the ground-state 
and the excited-state $Q \bar Q$ potentials, 
$\Delta E_n=V_n - V_0$.
The left panel shows no UV-cut case.
 The middle and right panels show the results with the UV-cut of 
 $\Lambda_{\rm UV}=3a_p \simeq 2.2{\rm GeV}$ and 
 $\Lambda_{\rm UV}=2a_p \simeq 1.5{\rm GeV}$, respectively. 
}
 \label{Fig:Ex_Energy}
\end{figure}

The gluonic excitation energy defined by the difference, 
$V_n - V_0$, is shown in Fig.\ref{Fig:Ex_Energy}. 
Roughly, even after the removal of UV-gluons, the magnitude of 
gluonic excitation is approximately unchanged, and 
the low-lying gluonic excitation remains to be of the order of 1GeV. 

\section{Summary and concluding remarks}

In this study, we have studied ground-state and low-lying even-parity 
excited-state potentials of quark-antiquark systems in terms of 
the gluon momentum component in the Coulomb gauge 
using SU(3) quenched lattice QCD.
By introducing UV-cut in the gluon momentum space, 
we study the ``UV-gluon sensitivity'' of 
the potentials and the stringy excitation.
Even after cutting off high-momentum gluon component above 1.5GeV, 
the IR part of the ground-state potential is almost unchanged. 
On the other hand, the change of excited-state potential 
is more significant by the cut of UV-gluons. 
However, the magnitude of the low-lying gluonic excitation 
remains to be of the order of 1GeV after the removal of UV-gluons.

As a next step, we will investigate 
the odd-parity excited-state potential, 
by using non-symmetric state operators of $\ket{\phi}_k$. 
In this work, we use the Coulomb gauge fixing, however,
to remove gauge artifact completely, 
it would be also interesting to apply 
the gauge-invariant expansion 
in terms of the Dirac mode \cite{Gongyo:2012}.

\section*{Acknowledgements}
  H.S. is supported by the Grant for Scientific Research 
  [(C) No.23540306, Priority Areas ``New Hadrons'' (E01:21105006)] 
  and T.I. is supported by a Grant-in-Aid for JSPS Fellows [No. 23-752] 
  from the Ministry of Education, Culture, Science and Technology (MEXT) 
  of Japan. 
  This work is supported by the Global COE Program, 
``The Next Generation of Physics, Spun from Universality and Emergence''.
  The lattice QCD calculation has been done on NEC-SX8R at Osaka University.

\end{document}